\documentclass[aps,preprint,showpacs]{revtex4}
\usepackage{bm}
\usepackage{amsmath}
\usepackage{graphicx}

\begin{document}

\title{Vortex stability in nearly two-dimensional Bose-Einstein condensates
with attraction}
\author{Dumitru Mihalache$^{1,2}$, Dumitru Mazilu$^{1,2}$, Boris A. Malomed$^{3}$,
and Falk Lederer$^{1}$,} \affiliation{$^{1}$ Institute of Solid
State Theory and Theoretical Optics, Friedrich-Schiller
Universit{\"a}t Jena, Max-Wien-Platz 1, D-077743 Jena, Germany}
\affiliation{$^{2}$National Institute of Physics and Nuclear
Engineering, Institute of Atomic Physics, Department of Theoretical
Physics, P.O. Box MG-6, Bucharest, Romania}
\affiliation{$^3$Department of Interdisciplinary Studies, Faculty of
Engineering, Tel Aviv University, Tel Aviv 69978, Israel}

\begin{abstract}
We perform accurate investigation of stability of localized vortices
in an effectively two-dimensional (``pancake-shaped") trapped BEC
with negative scattering length. The analysis combines computation
of the stability eigenvalues and direct simulations. The states with
vorticity $S=1$ are stable in a third of their existence region,
$0<N<(1/3)N_{\max }^{(S=1)}$, where $N$ is the number of atoms, and
$N_{\max }^{(S=1)}$ is the corresponding collapse threshold. Stable
vortices easily self-trap from arbitrary initial configurations with
embedded vorticity. In an adjacent interval, $(1/3)N_{\max
}^{(S=1)}<N<$ \ $\allowbreak 0.43N_{\max }^{(S=1)}$, the unstable
vortex periodically splits in two fragments and recombines. At $N>$
\ $\allowbreak 0.43N_{\max }^{(S=1)}$, the fragments do not
recombine, as each one collapses by itself. The results are compared
with those in the full 3D Gross-Pitaevskii equation. In a moderately
anisotropic 3D configuration, with the aspect ratio $\sqrt{10}$, the
stability interval of the $S=1$ vortices occupies $\approx 40\%$ of
their existence region, hence the 2D limit provides for a reasonable
approximation in this case. For the isotropic 3D configuration, the
stability interval expands to $65\%$ of the existence domain.
Overall, the vorticity heightens the actual collapse threshold by a
factor of up to $2$. All vortices with $S\geq 2$ are unstable.
\end{abstract}

\maketitle

\section{Introduction}

Vortices are fundamental dynamical objects in Bose-Einstein
condensates (BECs) \cite{Cornell}. They can be easily created in
large numbers, forming vortex lattices \cite{Engels}. A review of
basic results for the BEC vortices can be found in Ref.
\cite{vortex-review}. In most works, vortices were studied in
self-repulsive BECs, with a positive scattering length
characterizing collisions between atoms. In the experiment,
vortices are created stirring the condensate by a properly
designed laser beam \cite   {Cornell}, or by imprinting an
appropriate phase pattern onto a condensate trapped in a
Joffe-Pritchard magnetic trap \cite{Ketterle}. Recently, more
complex vortical structures were predicted in repulsive
condensates, such as vortex dipoles \cite{Lucian1-China} and
quadrupoles \cite{Lucian2}, and patterns in the form of globally
linked topological defects \cite{Lucian3}.

Vortices in repulsive BECs may be regarded as dark solitons on a
finite background, similar to optical vortices in a
self-defocusing dielectric medium \cite{Swartz}. A challenging
issue is the study of \textit{vortex solitons} (completely
localized objects with embedded vorticity) in attractive BECs,
with a negative scattering length \cite{Lewenst}. In particular,
attractive interaction occurs in the $^{7}$Li condensate \cite
{Hulet}, where quasi-one-dimensional solitons were created
\cite{Li}.

The most natural setting for vortex solitons is a ``pancake"
condensate, strongly confined in one direction ($z$) and weakly
trapped in the radial direction ($r$) in the transverse plane. In
the experiment, the ``pancake" is created by a superposition of a
tight optical trap, with a confinement thickness $a_{z}$, and a
loose radial magnetic trap with a trapping frequency $\Omega _{r}$
in the transverse plane \cite   {Ketterle2}. The theoretical
description of the ``pancake" configuration assumes that the
underlying three-dimensional (3D) Gross-Pitaevskii equation (GPE)
may be reduced to its 2D counterpart. The condition necessary for
the reduction is that the squared harmonic-oscillator length
determined by the radial trapping frequency $\Omega _{r}$,
\begin{equation}
\left( a_{r}^{(\mathrm{ho})}\right) ^{2}=\hbar /\left( m\Omega _{r}\right)
\label{ho}
\end{equation}
($m$ is the atomic mass), must be much larger than $a_{z}^{2}$, i.e.,
\begin{equation}
\Omega _{r}\ll \pi ^{2}\hbar /\left( ma_{z}^{2}\right) .  \label{pancake}
\end{equation}
For Li atoms trapped in the diffraction-limited gap between two
parallel strongly repulsive (blue-detuned) light sheets, with
$a_{z}\,\simeq \,2$ $\mu $m, this means $\Omega _{r}\ll 10$ KHz,
which is readily met in the experiment, where confining frequencies
are measured in tens of Hz. Condition (\ref{pancake}) also rule out
bending instability of vortex cores in the pancake geometry (in the
full 3D case, the instability results in decay of straight vortices
into vortex rings, in repulsive condensates \cite{vortex-to-ring}),
as $a_{z}$ is much smaller than the vertical length necessary for
the bending.

The condition (\ref{pancake}) for the applicability of the 2D
approximation can be cast in another form, which sets a limit on
the number of atoms $\mathcal{N}$ in the pancake-shaped
condensate. As shown below, this form is
\begin{equation}
\mathcal{N}~\,_{\sim }^{<}\,\frac{\hbar }{m\Omega _{r}|a|a_{z}},  \label{N}
\end{equation}
where $a$ is the (negative) scattering length. Taking the same value
$a_{z}=2 $ $\mu $m as above, together with experimentally relevant
$\Omega _{r}=10$ Hz [in this case, Eq. (\ref{ho}) yields
$a_{r}^{(\mathrm{ho})}\simeq 10$ $\mu $m for Li atoms], and $a=-0.1$
nm, which is typical for soliton experiments in $^{7}$Li
\cite{Hulet,Li}, Eq. (\ref{N}) yields $\mathcal{N}~\,_{\sim
}^{<}\,~10^{4}$.

Generally, vortex solitons are unstable against collapse
(catastrophic self-compression of the wave function), and also
against azimuthal perturbations which break the axial symmetry of
the solitons. In fact, condition (\ref{N}) provides for the
stability of the quasi-2D solitons against the 3D collapse. Within
the framework of the 2D description, both the ordinary
(zero-vorticity) solitons and their vortical counterparts can be
stabilized by a square-shaped optical lattice (OL)
\cite{BBB,Ziad}. The same OL can also support weakly localized
vortex solitons of the \textit{gap type} in a repulsive BEC
\cite{Estoril,gap-vortex}; a quasi-1D lattice stabilizes ordinary
solitons, but not vortices, in the 2D model with self-attraction
\cite{Estoril,quasi}. In attractive condensates, ordinary 3D
solitons can be stabilized by low-dimensional OLs [viz., a
quasi-2D lattice \cite{Estoril,quasi,PRE2004Mihalache}, or an
axially symmetric one \cite   {PRL2005Mihalache}], as well as by
the full 3D lattice \cite   {PRA2005Mihalache} (for a recent
review of multidimensional solitons in optics and BEC, see Ref.
\cite{JOB2005}). In addition, 2D zero-vorticity solitons can be
stabilized by means of the \textit{\ Feshbach-resonance
management} \cite{FRM}, which amounts to time-periodic modulation
of the scattering length between positive and negative values
\cite{1D}. However, the latter mechanism cannot stabilize vortex
solitons \cite{Isaac}, although it may inhibit the development of
their instability \cite{preprint}.

While vortex solitons of the 2D GPE with a negative scattering
length are strongly unstable in free space, they can be stabilized
by the radial trapping potential, $(m/2)\Omega _{r}^{2}r^{2}$. This
issue was considered in an early work \cite{Dodd} and more recently
\cite{Tristram,Carr}. It is well known that the radial trap
stabilizes zero-vorticity ($S=0$) states against collapse, provided
that the norm of the solution (which is proportional to the number
of atoms in the condensate) is limited from above. In the 3D case
with the isotropic radial trap, the maximum norm depends on the
corresponding trapping frequency $\Omega _{r}$. This maximum norm
was found by means of the variational approximation and from
numerical computations \cite{Dodd,Tristram}. In contrast to that,
the maximum 2D norm, $N_{\mathrm{\max }}^{(S=0)}$, \emph{does not}
depend on $\Omega _{r}$, being exactly equal to the norm of the
corresponding \textit{Townes soliton} in the free 2D space
\cite{Berge}. Similarly, the maximum value of the norm for the
trapped fundamental vortex (one with $S=1$), which does not lead to
collapse, depends on $\Omega _{r}$ in the 3D case (the shape of the
3D trap which maximizes the collapse threshold under additional
conditions was studied in Ref. \cite{Yuka}), but not in the 2D
situation, where it coincides with the known collapse threshold,
$N_{\mathrm{\max }}^{(S=1)}$, of the fundamental vortex soliton in
free space \cite{Kruglov}. Formally, the vorticity gives rise to a
dramatic increase of the 2D collapse threshold: $N_{\mathrm{\max
}}^{(S=1)}\approx 4N_{\mathrm{\max }}^{(S=0)}$. However, we will
demonstrate that the actual increase of the 2D stability threshold,
$N_{\mathrm{thr}}$, in the fundamental vortex is a more modest
effect, amounting to $N_{\mathrm{thr}}\simeq 2N_{\mathrm{\max
}}^{(S=0)}$; in the remaining part of the existence interval for the
vortex, $N_{\mathrm{ thr}}<N<N_{\mathrm{\max }}^{(S=1)}$, the
azimuthal instability splits the vortex into a pair of fragments,
which then collapse intrinsically.

It is relevant to mention that, in various models of nonlinear optics (see
Refs. \cite{FirthSkryabin} and brief reviews \cite{Pramana}), vortex
solitons, while being stable against radial perturbations, are easily split
into fragments by an azimuthal instability. Nevertheless, \emph{azimuthally
stable} vortical solitons have been predicted too, in models with competing
\cite{competing} or nonlocal nonlinearities \cite{nonlocal}, and in
defocusing media with an imprinted cylindrical lattice \cite{Yaroslav}.

In this work, we aim to investigate the stability of vortices in
the 2D GPE equation with self-attraction and weak radial trapping
potential, through computation of a full set of the corresponding
stability eigenvalues. The results will be verified by direct
simulations. We conclude that all vortices with $S\geq 2$ are
unstable (as conjectured in Ref. \cite{Tristram}), while the
fundamental ones ($S=1$) are completely stable for $0<N<N_{
\mathrm{cr}}\approx (1/3)N_{\mathrm{\max }}^{(S=1)}$ [in Ref.
\cite{Tristram}, it was conjectured that $N_{\mathrm{cr}}\simeq
(1/4)N_{\mathrm{\max }}^{(S=1)}$]. At $N=N_{\mathrm{cr}}$, the
$S=1$ vortex is destabilized by symmetry-breaking oscillatory
perturbations. Further, in an adjacent interval,
$(1/3)N_{\mathrm{\max }}^{(S=1)}<N<0.43N_{\mathrm{\max }}^{(S=1)}$
, the evolution of the unstable vortex \emph{does not } lead to
collapse; instead, it features nearly periodic splittings in two
fragments followed by their recombinations back into the vortex.
This regular dynamical regime was not known before. In the
remaining part of the existence interval, $0.43N_{\mathrm{\max
}}^{(S=1)}<N<N_{\mathrm{\max }}^{(S=1)}$, fragments produced by
the splitting of the vortex do not recombine, but rather blow up
intrinsically, as their individual norm exceeds the collapse
threshold for the Townes soliton.

We will also verify that the results obtained within the framework
of the 2D model are valid indeed in the full 3D GPE with the
corresponding anisotropic trap. In particular, in a moderately
anisotropic 3D model, with the radial-confinement length
$a_{r}^{(\mathrm{ho})}$ [see Eq. (\ref{ho})] larger by a factor of
$\sqrt{10}$ than its counterpart, $a_{z}^{(\mathrm{ho})}$, in the
orthogonal (tightly confined) direction, the relative size of the
stability area inside the existence region for 3D solitons with
the embedded vorticity $S=1$ is $\approx 0.4$, which is to be
compared to the above-mentioned relative size of the stability
region, $1/3$, in the 2D limit. Approaching the isotropic limit,
$a_{z}^{(\mathrm{ho})}/a_{r}^{(   \mathrm{ho})}\rightarrow 1$, the
stability area expands, attaining the relative size $\approx
\allowbreak 0.65$.

The paper is organized as follows. In Section 2, we formulate the 2D model,
briefly recapitulating its derivation from the 3D GPE. In Section 3, we
present basic results for the 2D model: shapes of the vortex states
(actually, we find them by continuation of exact states with a definite
value of the angular momentum in the linear 2D Schr\"{o}dinger equation),
and eigenvalues that determine their stability. In Section 4, we verify the
predicted stability conditions by direct simulations. In Section 5, we
consider the full 3D model with the anisotropic trap, and Section 5
concludes the paper.

\section{The model}

We assume a self-attractive condensate which is loaded in a nonrotating
loose trap in the $\left( x,y\right) $ plane, and tightly confined in the $z$
direction. The corresponding 3D GPE for the single-atom wave function $\Psi $
is \cite{book}:
\begin{equation}
i\hbar \frac{\partial \Psi }{\partial t}=\left\{ -\frac{\hbar
^{2}}{2m}\nabla ^{2}+\frac{m}{2}\left[ \Omega _{r}^{2}\left(
x^{2}+y^{2}\right) +\Omega _{z}^{2}z^{2}\right] +\frac{4\pi \hbar
^{2}a}{m}|\Psi |^{2}\right\} \Psi ,  \label{3DGP}
\end{equation}
where $\nabla ^{2}$ is the 3D Laplacian and $a$ is the negative
$s$-scattering length accounting for the attraction between atoms. In the
pancake configuration, the confinement frequency $\Omega _{z}$ is
assumed to be much larger than $\Omega _{r}$. If the tight
confinement in the $z$-direction is provided by parallel light
sheets repelling the atoms, the corresponding potential term,
instead of $m\Omega _{z}^{2}z^{2}/2$, is a deep rectangular
potential well in the region of $0<z<a_{z}$.

The 3D equation can be reduced to an effective GPE in two
dimensions, provided that the transverse quantum pressure is much
stronger than self-attraction. In other words, the energy of the
atom in the ground state of the vertical trap, if the latter is
taken as the deep potential well of width $a_{z}$,
\begin{equation}
E_{0}=\left( \pi \hbar \right) ^{2}/\left( 2ma_{z}^{2}\right) ,
\label{groundstate}
\end{equation}
must be much larger than the contribution of the attraction to the
atomic chemical potential, $\Delta \mu \sim -2\pi \hbar ^{2}\left(
a/m\right) n$, where $n\equiv \left\vert \Psi \right\vert ^{2}$ is
the 3D atom density [if the tight vertical confinement is provided
by large $\Omega _{z}$ in Eq. (\ref{3DGP}), then the respective
harmonic-oscillator length, $a_{z}^{(\mathrm{ho})}=\sqrt{\hbar
/\left( m\Omega _{z}\right) }$, taken as per Eq. (\ref{ho}), defines
the effective confinement size as $a_{z}=\pi a_{z}^{(
\mathrm{ho})}$].

For the pancake-shaped configuration in the weak radial trap, the density is
determined, together with a characteristic radial size $R$ of the pancake
itself, by the condition of the balance between the radial quantum pressure,
radial trapping force, and nonlinear self-attraction, which yields
\begin{equation}
R\sim \sqrt{\frac{\hbar }{m\Omega _{r}}}\equiv a_{r}^{(\mathrm{ho})},~n\sim
\frac{m\Omega _{r}}{8\pi a\hbar }~.  \label{2D}
\end{equation}
The substitution of expressions (\ref{groundstate}) and (\ref{2D}) in the
above condition $E_{0}\gg \left\vert \Delta \mu \right\vert $, which implies
the tight transverse confinement, leads to Eq. (\ref{pancake}) that
determines the range of the radial trapping frequencies where the 2D
description is applicable. As a consequence of this condition, the aspect
ratio of the pancake configuration is always large, $\pi R/a_{z}\sim
a_{z}^{-1}\sqrt{\hbar /\left( m\Omega _{r}\right) }\gg 1$. A limitation on
the number of atoms in the pancake state can be derived by noting that it is
determined by the product of the density, as given by Eq. (\ref{2D}), and
the effective volume, $\simeq \pi R^{2}a_{z}$, which leads to Eq. (\ref{N}).
If $\mathcal{N}$ exceeds this limit, the condensate will start a nearly-2D
collapse, which may later go over into a stronger collapse in three
dimensions.

The final derivation of the effective 2D equation follows a well-known
route: the 3D wave function is factorized,
\begin{equation}
\Psi (x,y,z,t)=\psi (x,y,t)\phi _{0}(z),  \label{Psipsi}
\end{equation}
where $\phi _{0}=\sin \left( \pi z/a_{z}\right) $ is the ground-state wave
function of the transverse state between two hard walls separated by the
distance $a_{z}$. After that, averaging of the 3D equation (\ref{3DGP}) in
the $z$ direction leads to a normalized two-dimensional GPE \cite{2D},
\begin{equation}
i\frac{\partial \psi }{\partial t}=\left[ -\frac{1}{2}\left(
\frac{\partial ^{2}}{\partial x^{2}}+\frac{\partial ^{2}}{\partial
y^{2}}\right) +\frac{1}{2}\Omega _{r}^{2}\left( x^{2}+y^{2}\right)
+g|\psi |^{2}\right] \psi , \label{GP}
\end{equation}
where $g\equiv 3\pi a$ (or $g\equiv 2\sqrt{2}\pi a$ for the case of the
parabolic trap in the vertical direction), the transverse-ground-state
energy $E_{0}$ was subtracted from the chemical potential, while $\Omega _{r}
$ and $t$ actually stand for $m\Omega _{r}/\hbar $ and $\left( \hbar
/m\right) t$. The variables $x$, $y$ and $|\psi |^{2}$ in Eq. (\ref{GP}) are
measured in the same units as in Eq. (\ref{3DGP}). We then redefine them
too, by $\psi \rightarrow \sqrt{|g|/\Omega _{r}}\psi $, $\left( x,y\right)
\rightarrow \sqrt{\Omega _{r}}\left( x,y\right) $, and additionally rescale
time by $t\rightarrow \Omega _{r}t$, to finally set $\Omega _{r}=-g\equiv 1$
(unless $g=0$).

Equation (\ref{GP}) conserves the 2D norm, $N~=~\int \int \left\vert \psi
(x,y,t)\right\vert ^{2}dxdy$, and energy
\begin{equation}
E=\int \int \left[ \frac{1}{2}\left( \left\vert \frac{\partial
\psi }{\partial x}\right\vert ^{2}+\left\vert \frac{\partial \psi
}{\partial y}\right\vert ^{2}\right) -\frac{1}{2}g|\psi
|^{4}+\frac{1}{2}\Omega _{r}^{2}(x^{2}+y^{2})|\psi |^{2}\right]
dxdy.  \label{E}
\end{equation}
A straightforward consequence of Eq. (\ref{GP}) is the relationship $E=\mu
N+\pi |g|\int_{0}^{\infty }rR^{4}dr$, which is used to cross-check the
accuracy of numerical results, see below. Further, the above rescalings and
factorization (\ref{Psipsi}) result in the following relation between the 2D
norm and the number of atoms, $\mathcal{N}$ (which is defined as the norm of
the 3D wave function in original units),
\begin{equation}
\mathcal{N}=\frac{a_{z}}{2|g|}N\equiv \frac{a_{z}}{6\pi |a|}N,  \label{NN}
\end{equation}
where the second equality follows from the above relation $g=3\pi a$, if the
vertical confinement is provided by the repelling light sheets.
If, instead, a tight parabolic potential is used to confine the
condensate in the $z$-direction, the coefficient in front of the
last term in Eq. (\ref{NN}) is replaced by
$a_{z}^{(\mathrm{ho})}/\left( 2\sqrt{2}|a|\right) $, where $
a_{z}^{(\mathrm{ho})}$ is the respective linear-oscillator length
as defined above.

The final normalized form of the 3D GPE, which is also considered below, is
\begin{equation}
i\frac{\partial \Psi }{\partial t}=\left[ -\frac{1}{2}\left( \frac{\partial
^{2}}{\partial x^{2}}+\frac{\partial ^{2}}{\partial y^{2}}+\frac{\partial
^{2}}{\partial z^{2}}\right) +\frac{1}{2}\left( x^{2}+y^{2}+\Omega
^{2}z^{2}\right) -|\Psi |^{2}\right] \Psi .  \label{3D}
\end{equation}
Here the nonlinearity coefficient is again scaled to be $-1$, and $\Omega
^{2}\equiv \Omega _{z}^{2}/\Omega _{r}^{2}$ measures the relative strength
of the tight confinement in the vertical direction, the nearly-2D case
corresponding to $\Omega ^{2}\gg 1$. The norm of the 3D solutions, $N=\int
\int \int \left\vert \Psi (x,y,z,t)\right\vert ^{2}dxdydz$, is related to
the number of atoms as follows:
\begin{equation}
\mathcal{N}=\sqrt{\frac{\hbar }{m\Omega _{r}}}\frac{N}{8\pi
|a|}\equiv \frac{a_{r}^{(\mathrm{ho})}}{8\pi |a|}N, \label{N3D}
\end{equation}
where $a_{r}^{(\mathrm{ho})}$ is the harmonic-oscillator length in the
radial direction, defined as per Eq. (\ref{2D}). The conserved energy of the
3D condensate is
\begin{equation}
E=\int \int \int \left[ \frac{1}{2}\left( \left\vert
\frac{\partial \Psi }{\partial x}\right\vert ^{2}+\left\vert
\frac{\partial \Psi }{\partial y}\right\vert ^{2}+\left\vert
\frac{\partial \Psi }{\partial z}\right\vert ^{2}\right)
-\frac{1}{2}|\Psi |^{4}+\frac{1}{2}(x^{2}+y^{2}+\Omega
^{2}z^{2})|\Psi |^{2}\right] dxdydz.  \label{E3D}
\end{equation}
Note that, after $\Omega _{r}$ and $-g$ were normalized to be $1$, the
two-dimensional GPE (\ref{GP}) features no free parameters, while its 3D
counterpart, Eq. (\ref{3D}), contains one free coefficient, $\Omega ^{2}$.

\section{Two-dimensional vortex solutions and their linear stability}

Stationary vortex solutions to Eq. (\ref{GP}) are looked for as
\begin{equation}
\psi =R(r)\exp \left( iS\theta -i\mu t\right) ,  \label{psiR}
\end{equation}
with an integer vorticity $S$, real chemical potential $\mu $,
and function $R(r)$ obeying the equation
\begin{equation}
R^{\prime \prime }+r^{-1}R^{\prime }+\left( 2\mu -S^{2}r^{-2}-r^{2}\right)
R-2gR^{3}=0,  \label{Rparabolic}
\end{equation}
where the prime stands for $d/dr$. In the linear case, $g=0$, Eq.
(\ref{Rparabolic}) is tantamount to the usual stationary
quantum-mechanical Schr\"{o}dinger equation for the 2D harmonic
oscillator, which gives rise to an infinite set of discrete
eigenvalues, $\mu =j+k+1$, with $j,k=0,1,2,...$ . In the Cartesian
coordinates, the corresponding eigenfunctions are (recall we have
set $\Omega \equiv 1$)
\begin{equation}
\psi _{jk}(x,y,t)=e^{-i\left( j+k+1\right) t}\Phi _{j}(x)\Phi _{k}(y),
\label{xy}
\end{equation}
where $\Phi _{j}(x)$ and $\Phi _{k}(x)$ are stationary wave functions of the
1D harmonic oscillator associated with the energy eigenvalues
$j+1/2$ and $k+1/2$, respectively. The set of states (\ref{xy}) is
degenerate, as all the states corresponding to a fixed value of
$j+k$ appertain to the same eigenvalue. Making use of this
degeneracy, the eigenfunctions may be rearranged into combinations
with a definite value of the vorticity, which is
\begin{equation}
S=j+k\equiv \mu -1.  \label{Smu}
\end{equation}
In particular, the combination
\begin{equation}
\psi _{0}^{(S=1)}\equiv \psi _{10}(x,y,t)+i\psi _{01}(x,y,t)\equiv r\exp
\left( -2it+i\theta -r^{2}/2\right)   \label{S=1}
\end{equation}
is the wave function with the lowest value of the chemical potential for $S=1
$, $\mu \equiv j+k+1=2$, and
\begin{equation}
\psi _{0}^{(S=2)}\equiv \psi _{20}(x,y,t)-\psi _{02}(x,y,t)+2i\psi
_{11}(x,y,t)\equiv r^{2}\exp \left( -3it+2i\theta -r^{2}/2\right)
\label{S=2}
\end{equation}
is the wave function with the lowest chemical potential, $\mu
\equiv j+k+1=3$, for $S=2$.

Nonlinear solutions for vortices can be obtained by a continuation
method, starting at $g=0$ with expressions (\ref{S=1}) and
(\ref{S=2}). Note that, pursuant to Eq. (\ref{Rparabolic}), the
shift of $\mu $ due to the self-attractive nonlinearity (with $g<0$)
is negative, hence from the comparison with the linear result
(\ref{Smu}) it follows that
\begin{equation}
\mu (N\neq 0)<\mu (N=0)\equiv \mu _{\mathrm{\max }}=S+1.  \label{cutoff}
\end{equation}

To specify solutions to Eq. (\ref{Rparabolic}) in the nonlinear
case (recall we set $g=-1$), Eq. (\ref{Rparabolic}) was solved
numerically. The dependences $\mu (N)$ obtained from the numerical
solutions for $S=0,1$ and $2$ are plotted in Fig. 1(a) (in a part,
these dependences are tantamount to those reported earlier in Ref.
\cite{Tristram}). Naturally, $\mu $ approaches the linear value
$\mu _{\mathrm{\max }}$ for $N\rightarrow 0$, see Eq.
(\ref{cutoff}), while for large negative $\mu $ the norm of the
$S=0$ state asymptotically approaches the value $N_{\mathrm{\max
}}^{(S=0)}\approx 5.85$, which is the above-mentioned collapse
threshold corresponding to the Townes soliton \cite{Berge}.
Similarly, for the states with $S\geq 1$ the asymptotic values
$N(\mu =-\infty )$ are the above-mentioned ones, $N_{\max
}^{(S)}$, which determine the (formal) collapse threshold for the
vortices.

To study stability of the stationary solutions, we search for a perturbed
solution of Eq. (\ref{GP}) as
\begin{equation}
\psi (x,y,t)=[R(r)+u(r)\exp (\lambda t+iL\theta )+v^{\ast }(r)\exp (\lambda
^{\ast }t-iL\theta )]\exp \left( iS\theta -i\mu t\right) ,  \label{pert}
\end{equation}
where $\left( u,v\right) $ and $\lambda $ are eigenmodes and the instability
growth rate corresponding to an integer azimuthal index $L$ of the
perturbation. The linearization around the stationary solution leads to
equations
\begin{eqnarray}
i\lambda u+\frac{1}{2}\left[ u^{\prime \prime }+r^{-1}u^{\prime
}-(S+L)^{2}r^{-2}u\right] +\mu u+R^{2}(v+2u)-\frac{1}{2}r^{2}u &=&0,  \notag
\\
-i\lambda v+\frac{1}{2}\left[ v^{\prime \prime }+r^{-1}v^{\prime
}-(S-L)^{2}r^{-2}v\right] +\mu v+R^{2}(u+2v)-\frac{1}{2}r^{2}v &=&0,
\label{growth}
\end{eqnarray}
which were solved numerically, with boundary conditions demanding that $u(r)$
and $v(r)$ decay exponentially at $r\rightarrow \infty $, and
decay as $r^{\left\vert S\pm L\right\vert }$ at $r\rightarrow 0$.

The most important result following from the numerical computation of
eigenvalues $\lambda $ is that the vortex soliton with $S=1$ is stable,
i.e., real parts of all the eigenvalues remain equal to zero, only in a
\emph{finite interval} of values of the chemical potential adjacent to the
linear limit [see Eq. (\ref{cutoff})], $\mu _{\mathrm{\max }}(S=1)\equiv
2>\mu >\mu _{\mathrm{cr}}\approx 1.276$, and outside this interval, i.e.,
for $-\infty <\mu <\mu _{\mathrm{cr}}$, the states with $S=1$ are unstable.
At the critical point ($\mu =\mu _{\mathrm{cr}}$), the corresponding
critical norm is $N_{\mathrm{cr}}^{(\mathrm{2D})}=7.79$. Thus, the vortices
with $S=1$ are stable in the region
\begin{equation}
0\leq N<N_{\mathrm{cr}}^{(2D)}=7.79,  \label{main}
\end{equation}
and in the remaining \emph{two-thirds} of their existence interval, i.e.,
\begin{equation}
7.79<N<N_{\mathrm{\max }}^{(S=1)}\approx 24.1,  \label{unstable}
\end{equation}
they are \emph{unstable}. The conclusion that ``less nonlinear"
solutions, i.e., ones with smaller $N$, are more stable is quite
natural, as they continue the stable linear solutions which
correspond to $N=0$ \cite{Ulf}.

The energy [see Eq. (\ref{E})] at $\mu =\mu _{\mathrm{cr}}$ is
$E_{\mathrm{cr }}=12.913$. The energy as a function of $N$ is
displayed, for the $S=1$ soliton family, in Fig. 1(b). It is
noteworthy that, while the $\mu (N)$ dependence in Fig. 1(a) is
monotonic, its $E(N)$ counterpart is not.

To further characterize the (in)stability of the vortices with $S=1$
and $2$, in Fig. 2(a) we show the largest real part of the
eigenvalues $\lambda _{L}^{(S)}$ as a function of $\mu $ for
different values of the perturbation azimuthal index $L$, see Eq.
(\ref{pert}). Extensive numerical calculations demonstrate that the
perturbation mode that actually determines the stability border
always has $L=2$. In particular, $\mathrm{Re}\left( \lambda
_{L=2}^{(S=2)}\right) $ is positive in the entire existence domain
of the $S=2$ vortices, up to the linear limit, $\mu _{\max }(S=2)=3$
[see Eq. (\ref{cutoff})], thus making the vortices with $S=2$
\emph{completely unstable} and confirming a conjecture put forward
in Ref. \cite{Tristram}. All vortices with $S>2$ are unstable too.

For the fundamental vortex solitons with $S=1$, the numerical
calculation of the eigenvalues $\lambda _{L}^{(S=1)}$ was
performed for values of the azimuthal index up to $L=6$, in a
broad range of values of the soliton's chemical potential,
$-20<\mu <2$. It was found that the maximum value of $
\mathrm{Re}\left( \lambda _{L}^{(S=1)}\right) $ is nonzero for
$L=1,2,3$, and this value is zero (up to the numerical accuracy)
for higher values of the azimuthal index, $L=4,5$, and $6$. To
further check the accuracy of the numerical results, we have
compared the eigenvalues produced by the numerical code running on
400 and 800 discretization points, respectively, and concluded
that four significant digits are identical in both cases.

The bifurcation responsible for the destabilization of the $S=1$
vortex at $\mu =\mu _{\mathrm{cr}}$ is identified by results
displayed in Fig. 2(b). As seen from the figure, the bifurcation
is of the \textit{Hamiltonian-Hopf} type \cite{HH}, i.e., it is
accounted for by a quartet of complex eigenvalues, $(\lambda
,\lambda ^{\ast },-\lambda ,-\lambda ^{\ast })$, cf. earlier
reported results for solitons in scalar \cite{Pego2002,osc1} and
vectorial \cite{osc2,osc3} models. The unstable quartet is
generated by a collision of two pairs of imaginary (stable)
eigenvalues, which is typical to the Hamiltonian-Hopf bifurcation.

\section{Direct simulations of two-dimensional stable and unstable vortices}

To verify the above predictions concerning the stability of the
vortex solitons in the 2D approximation, we have performed direct
simulations of the perturbed evolution of vortex solitons in Eq.
(\ref{GP}). The simulations were carried out by means of the
Crank-Nicholson discretization scheme. The system of the
corresponding nonlinear finite-difference equations was first
solved by means of the Picard iteration method \cite   {Ortega},
and the resulting linear system was then handled using the
Gauss-Seidel iterative scheme. For good convergence we typically
needed five Picard iterations and eight Gauss-Seidel iterations.
We employed a grid with $600\times 600$ points (typical stepsizes
were $\Delta x=\Delta y=0.025$ and $\Delta t=0.0008$).

To test the stability of the $S=1$ vortex, random noise was added to it at
the initial moment -- typically, with a $5\%$ relative amplitude. Figure 3
shows how a vortex with $S=1$, that was predicted above to be stable against
small perturbations, completely recovers after the addition of the initial
perturbations.

To further explore the robustness of the vortex which is predicted
to be stable, in Fig. 4 we display its self-trapping from an
arbitrarily chosen Gaussian input with a nested vortex,
$u_{0}=Ar\exp \left( -\alpha r^{2}+i\theta \right) $, that has
$A=1$ and $\alpha =1$. The shape of the input is far from the
exact stable vortex soliton. In this case, we used a grid with
$391\times 391$ points and stepsizes $\Delta x=\Delta y=0.028$, $
\Delta t=0.001$. Note that, at the values of $N$ corresponding to
the definitely stable solitons displayed in Figs. 3 and 4, a
rather inaccurate stability condition proposed in Ref.
\cite{Tristram} for the $S=1$ vortices would imply instability.

For the $S=1$ vortices which are predicted above to be unstable, we have
found that, in the interval of
\begin{equation}
7.79<N<10.30,  \label{intermediate}
\end{equation}
which is adjacent to stability region (\ref{main}) and occupies $\approx 1/8$
of the entire existence region of the vortices [cf. Eq. (\ref{unstable})],
initial perturbations initiate \emph{regular} quasi-periodic evolution,
which features recurrent splittings of the soliton into two segments and
their recombinations. The regular character of the evolution in this case is
also manifest in nearly periodic oscillations of the soliton's amplitude,
see typical examples in Figs. 5 and 6. This dynamical regime (which, to the
best of our knowledge, has never been reported before) can be easily
understood, as, unlike the known scenario of the instability development for
vortex solitons in nonlinear-optical models in free space, where the
splinters emerging after the breakup of the vortex soliton separate, moving
away in tangential directions, in the present setting they cannot do it,
being confined by the parabolic trap. We stress that Eq. (\ref{intermediate}
) shows that the introduction of the vorticity makes it possible to \emph{\
heighten} the actual collapse threshold for solitons in the trapped
self-attractive condensate by a factor of two: from $N_{\max }^{(S=0)}=5.85$
for the spinless solitons to $N\approx 11$ for their vortical counterparts.

For larger values of the norm, $10.3<N<N_{\mathrm{\max
}}^{(S=1)}\approx 24.1 $, the evolution of the unstable vortex
again starts with its splitting into two fragments; however, in
this case they do not recombine, but rather quickly blow up (not
shown here). The fact that the corresponding collapse threshold
for the $S=1$ vortex solitons, found at $N=10.3$, is much smaller
than the above-mentioned formal threshold, $N_{\mathrm{\max }
}^{(S=1)}\approx 24.1$, can be readily explained. Indeed, it has
been shown above that the azimuthal instability sets in earlier
than the radial instability that would directly lead to collapse.
If the norm of each of the zero-vorticity fragments, into which
the vortex is broken by the azimuthal instability, is close enough
to the collapse threshold for the $S=0$ solitons, i.e., $N_{\max
}^{(S=0)}=5.85$, the fragments start collapsing and therefore fail
to recombine into the vortex, unlike the situation with the weak
instability in interval (\ref{intermediate}). This explanation is
consistent with the fact that the actual collapse threshold for
the vortex, $N=10.3$ , is fairly close to $2N_{\max }^{(S=0)}$.

As concerns the $S=2$ solitons which were predicted above to be
unstable at all finite values of $N$, the simulations (not shown
here) demonstrate their irregular (chaotic) evolution for
$0<N<10.4$. For $N>10.4$, they exhibit collapse, again through the
splitting into a set of two $S=0$ solitons which then blow up
intrinsically. Lastly, we have also checked that the
zero-vorticity solitons in Eq. (\ref{GP}) exhibit collapse exactly
when $N>N_{\mathrm{\max }}^{(S=0)}=5.85$, as was assumed above.

At a late stage of the collapse, when the condensate will shrink
to a size $\sim a_{z}$, the above derivation of the 2D equation
(\ref{GP}) from its 3D counterpart (\ref{3DGP}) will break down,
and the initially quasi-2D collapse will switch into a faster 3D
blow-up mode. Direct simulations of the full 3D equation
(\ref{3D}) corroborate this expectation, as will be reported in
detail elsewhere.

\section{Vortex solitons in the three-dimensional Gross-Pitaevskii equation}

In this section, we aim to compare the above results, obtained in the
framework of the 2D approximation, with what can be found from the full 3D
equation (\ref{3D}), where the quasi-2D case corresponds to large $\Omega $.
Systematic presentation of the results for the 3D model will be a subject of
a separate work, while here we report, in a brief form, findings which are
most relevant for the comparison with the 2D approximation.

Stationary solutions for the 3D solitons with embedded vorticity
$S$ and chemical potential $\mu $ are looked for in the form [cf.
Eq. (\ref{psiR})] $\Psi =R(r,z)\exp \left( iS\theta -i\mu t\right)
$, where $r,z,\theta $ are cylindrical coordinates, and the real
function $R$ obeys the equation
\begin{equation}
\frac{\partial ^{2}R}{\partial r^{2}}+\frac{1}{r}\frac{\partial
R}{\partial r}+\frac{\partial ^{2}R}{\partial z^{2}}+\left( 2\mu
-\frac{S^{2}}{r^{2}}-r^{2}-\Omega ^{2}z^{2}\right) R-2R^{3}=0,
\label{3DR}
\end{equation}
supplemented by obvious boundary conditions: $R(r,z)\rightarrow 0$
when $r\rightarrow \infty $ or $|z|\rightarrow \infty $, and $R\sim r^{S}$
for $r\rightarrow 0$ at fixed $z$ (we again define $S$ to be $\geq 0$). After
finding a family of the vortices from numerical solution of Eq. (\ref{3DR}),
their stability was investigated via the computation of eigenvalues for
small perturbations.

First of all, the linear version of Eq. (\ref{3DR}) can be considered
similar to how it was done above for the 2D case: the solutions are built as
linear combinations of products of the wave functions of the harmonic
oscillator, cf. Eq. (\ref{xy}):
\begin{equation}
\Psi _{jkl}(x,y,z,t)=e^{-i\left[ j+k+1+\left( l+1/2\right) \Omega \right]
t}\Phi _{j}(x)\Phi _{k}(y)\Phi _{l}(\sqrt{\Omega }z),  \label{xyz}
\end{equation}
where $l=0,1,2,...$ is the quantum number in the $z$-direction.
Being interested here in the ``flattest" states (ones closest to the
pancake configuration in the 3D space), we will set $l=0$. Then,
wave functions with given vorticity $S$ in the $\left( x,y\right) $
plane are constructed as combinations of factorized wave functions
(\ref{xyz}), subject to the same constraint as in the 2D case, i.e.,
$j+k=S$. From this restriction and from the form of the chemical
potential in the linear-limit solution (\ref{xyz}) with $l=0$, $\mu
=j+k+1+\Omega /2$, it follows that the chemical potential of
solutions to the full nonlinear equation (\ref{3DR}), obtained as a
continuation starting from the linear limit, obeys a limitation
\begin{equation}
\mu <\mu _{\max }=S+1+\Omega /2\text{,}  \label{3Dcutoff}
\end{equation}
cf. Eq. (\ref{cutoff}).

Numerically found families of the 3D soliton solutions with $S=1$
are characterized by $\mu (N)$ and $E(N)$ curves displayed in Fig.
\ref{fig7}, one for the anisotropic model, with $\Omega =10$, and
one for its isotropic counterpart, with $\Omega =1$. In the former
case [which, as a matter of fact, is not a strongly anisotropic one,
as the corresponding aspect ratio of the confinement lengths in the
$\left( x,y\right) $ and $z$ directions is $\sqrt{10}$, pursuant to
Eq. (\ref{ho})], the existence and stability intervals of the
fundamental vortices have been found to be, respectively,
$0<N<14.46$, and
\begin{equation}
0<N<N_{\mathrm{cr}}^{(3D)}|_{\Omega =10}=5.95  \label{Omega=10}
\end{equation}
[cf. Eqs. (\ref{main}) and (\ref{unstable}) in the 2D model], hence the
relative size of the stability area is $\allowbreak \approx 0.41$.
This should be compared to the relative size of the stability area
in the 2D limit reported above, which is $\approx 1/3$. For the
isotropic model ($\Omega =1$), the existence and stability
intervals are, respectively, $0<N<23 $, and
\begin{equation}
0<N<N_{\mathrm{cr}}^{(3D)}|_{\Omega =1}\approx 15,  \label{Omega=1}
\end{equation}
their ratio being $\approx \allowbreak 0.65$.

Thus, we conclude that the 2D limit yields a reasonable
approximation for the pancake-shaped configurations (the relative
difference between its prediction and the direct numerical result
for the moderately anisotropic model with the aspect ratio
$\sqrt{10}$ is $\simeq 20\%$). Another noteworthy conclusion is that
the relative size of the stability area essentially increases as the
3D model approaches the isotropic configuration; in the isotropic
limit, the relative size of the stability region is almost twice as
large as in the 2D limit. In terms of the maximum value of the norm,
$N_{\max }$, up to which the vortex soliton remains stable, the
difference is still bigger: $N_{\max }$ for $\Omega =1$ exceeds its
counterpart for $\Omega =10$ by a factor of $\approx 2.5$.

To further illustrate the effect of the degree of asymmetry on the stability
of vortex solitons, in Fig. \ref{fig8} we display the spectrum of unstable
eigenvalues for the 3D solitons with $S=1$, in the same two cases as
considered above, i.e., the moderately anisotropic one with $\Omega =10$,
and isotropic with $\Omega =1$. Comparison of this figure with the
respective picture from Fig. \ref{fig2}(a) for the 2D model shows a gradual
deformation of the spectrum with the transition from the 2D limit to the
isotropic 3D configuration. An important observation is that the transition
from the 2D approximation to the full 3D model does not lead to any new
instability eigenmode (at least, up to the isotropic limit, $\Omega =1$);
incidentally, the bending instability of the vortex core does not occur
either.

As well as in the 2D case, vortex solitons with $S\geq 2$ [at
least, the ``flattest" ones, corresponding to $l=0$ in the linear
limit, see Eq. (\ref{xyz})] are all unstable. As concerns the
fundamental solitons with $S=0$, we could easily reproduce a known
result stating that, as well as in the 2D model, they are stable
within their full existence region, which is $N<3.8$ for $\Omega
=10$, and $N<7.3$ for $\Omega =1$. Comparing these values with
ones in Eqs. (\ref{Omega=10}) and (\ref{Omega=1}), we conclude
that the introduction of the vorticity enhances the stability
limit by a factor of $\allowbreak 1.6$ for $\Omega =10$, and by
$2$ for $\Omega =1$. Recall that, in the 2D model, the same
enhancement factor was $1.3$ [if its definition does not include
the extra region (\ref{intermediate}) of the quasi-stable
dynamical states]; this again demonstrates a general trend to the
expansion of stability regions with the transition from the 2D
limit to the fully isotropic 3D case.

\section{Conclusion}

We have reported results of accurate investigation of the stability of
localized vortices in Bose-Einstein condensates with self-attraction,
trapped in the nearly two-dimensional configuration. We have briefly
recapitulated the derivation of the effective 2D Gross-Pitaevskii equation
from its full 3D counterpart, and demonstrated that, in the 2D limit, the
vortices with $S=1$ are stable against azimuthal splitting in a \emph{third}
of their existence region, i.e., for $0<N<(1/3)N_{\max }^{(S=1)}$, in terms
of the solution's norm (which is proportional to the number of atoms trapped
in the vortex soliton). In the adjacent interval, $(1/3)N_{\max }^{(S=1)}<N<$
\ $\allowbreak 0.43N_{\max }^{(S=1)}$, the unstable vortex does not
collapse, but rather demonstrates stable quasi-periodic dynamics, featuring
cycles of splitting into two fragments and recombination, which is a novel
dynamical regime. For $N>0.43N_{\max }^{(S=1)}$, the two segments produced
by the splitting fail to recombine, as they collapse intrinsically. The
stable vortex soliton created in the pancake-shaped condensate of $^{7}$Li
may contain up to $10,000$ atoms.

The results were checked against direct simulations of the 2D
Gross-Pitaevskii equation, and, which is crucially important for the
verification of their physical relevance, they were compared with the
stability analysis of localized vortices in the full 3D equation. It was
concluded that, for the moderately anisotropic 3D configuration with the
aspect ratio $\sqrt{10}\approx \allowbreak 3.2$, the stability interval of
the $S=1$ vortices occupies $\approx 40\%$ of their existence region (hence,
the 2D limit provides for quite a reasonable approximation). In the 3D
isotropic limit, the stability interval expands to $65\%$ of the existence
domain. The enhancement factor for the collapse threshold, caused by the
introduction of vorticity $S=1$, was found too; in the 3D isotropic
configuration it attains the value of $2$. All higher-order localized
vortices, with $S\geq 2$, are completely unstable, in the 2D limit and in
the full 3D Gross-Pitaevskii equation alike.

Support from Deutsche Forschungsgemeinschaft (DFG), Bonn, is acknowledged.
The work of B.A.M. was partially supported by the Israel Science Foundation
through the Center-of-Excellence grant No. 8006/03.

\newpage

\begin{figure}[t]
\caption{Chemical potential $\protect\mu $ (a) and energy $E$ (b) versus
normalized number of atoms $N$ for solutions with vorticities $S=0,1,$ and $2
$, as found from Eq. (\protect\ref{Rparabolic}) with $g\equiv -1$. Solid and
dashed lines show stable and unstable branches of the solutions (see below).
The arrow indicates the point where the $S=1$ vortices loose their stability.
}
\label{fig1}
\end{figure}

\begin{figure}[t]
\caption{(a) The real part of the perturbation growth rate
$\protect\lambda $ for different azimuthal indices $L$. (b) The
real and imaginary parts (solid and dashed lines, respectively) of
the eigenvalues responsible for the \emph{\ Hamiltonian-Hopf}
bifurcation which destabilizes the vortices with $S=1$ at the
critical point, $\protect\mu =\protect\mu _{\mathrm{cr}}\approx
1.276$. This value and the bifurcation point are marked by
arrows.} \label{fig2}
\end{figure}

\begin{figure}[t]
\caption{Grey-scale plots illustrating recovery of the perturbed stable
vortex with $S=1$ for $\protect\mu =1.4$. (a,b): Intensity and phase
distributions in the initial configuration including random noise. (c,d):
The same in the self-cleaned vortex soliton at $t=120$. The norms of the
input and output solitons are $N=6.641$ and $N=6.629$, respectively (the
small loss of the norm is due to absorbers placed at borders of the
integration domain).}
\label{fig3}
\end{figure}

\begin{figure}[t]
\caption{Formation of a stable vortex with $S=1$ from a Gaussian
input. (a,b): The intensity and phase distributions in the initial
Gaussian with a nested vortex. (c,d): The intensity and phase
distributions in the output ($t=700$). The input and output
configurations have the norm, respectively, $N=6.406$ and
$N=6.392$.} \label{fig4}
\end{figure}

\begin{figure}[t]
\caption{Quasi-periodic evolution of the amplitude of the unstable $S=1$
vortex for $\protect\mu =1.2$. The labels a, b, c and d point at values of
the amplitude at $t=0,100,140$, and $180$ (see Fig. 6).}
\label{fig5}
\end{figure}

\begin{figure}[t]
\caption{Regular evolution of the unstable vortex with $S=1$ for
$\protect   \mu =1.2$ initiated by random noise, which shows
periodic splitting into two fragments and their subsequent
recombination: (a) $t=0$, (b) $t=100$, (c) $t=140$, and (d)
$t=180$. Here, values of the input and output norm are $ N=8.476$
(at $t=0$) and $N=8.460$ (at $t=240$).} \label{fig6}
\end{figure}

\begin{figure}[t]
\caption{The chemical potential (a) and energy (b) for families of
three-dimensional solitons with the embedded vorticity $S=1$ vs. the
normalized number of atoms, with the moderately anisotropic ($\Omega =10$)
and isotropic ($\Omega =1$) confining potentials. The continuous and dashed
parts of the curves show, respectively, stable and unstable parts of the
families. }
\label{fig7}
\end{figure}

\begin{figure}[t]
\caption{The same as in Fig. \protect\ref{fig2}(a), but for vortex
solitons with $S=1$ in the full three-dimensional model: (a)
$\Omega =10$; (b) $\Omega =1$. The arrows show the edge of the
existence regions of the solitons on the scale of the chemical
potential, as per Eq. (\protect\ref{3Dcutoff}).} \label{fig8}
\end{figure}

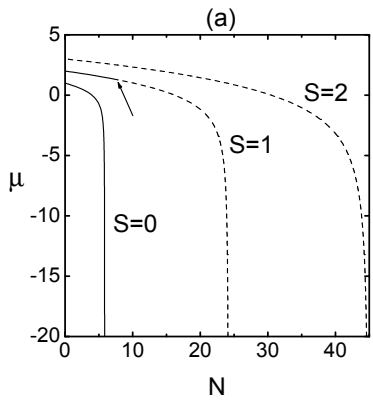
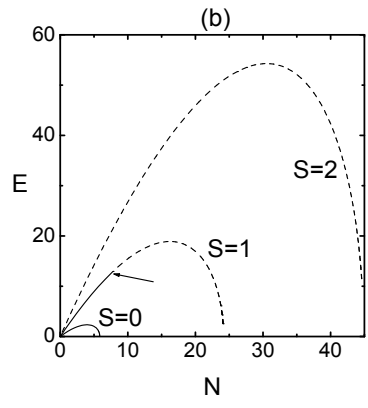
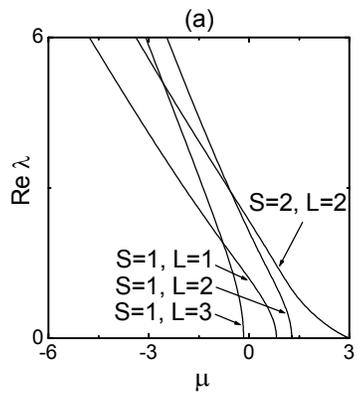
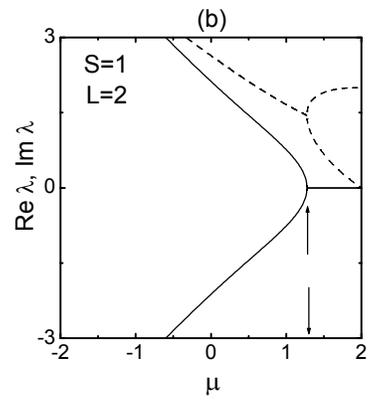
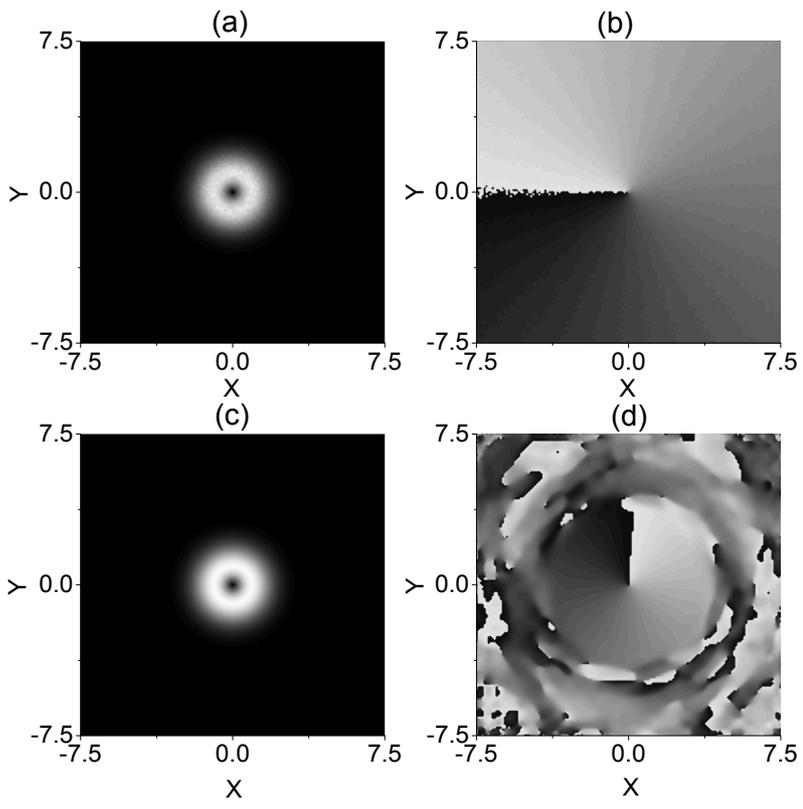
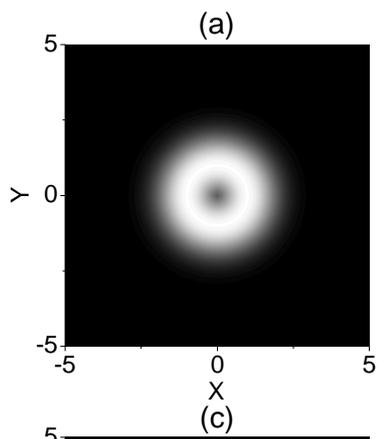
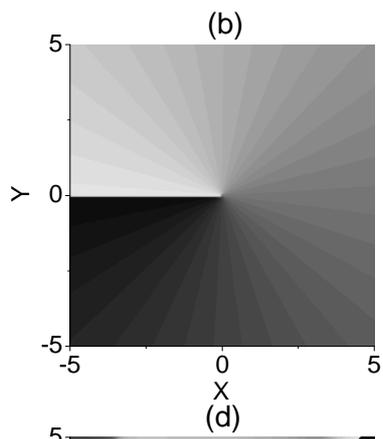
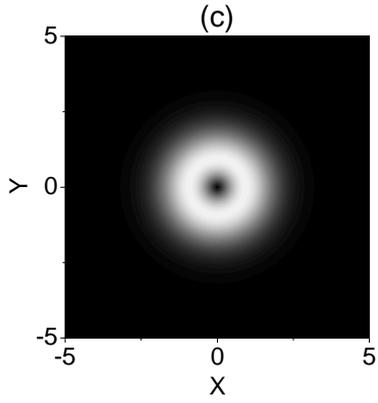
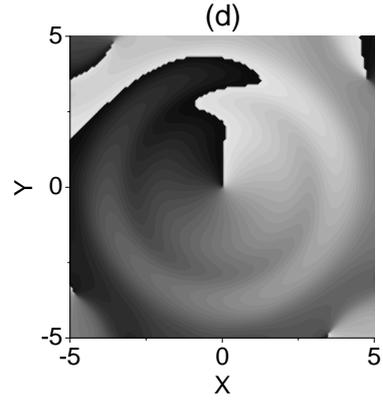
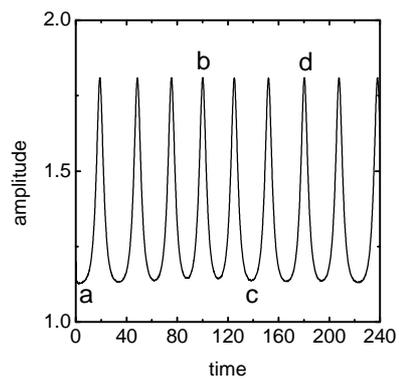
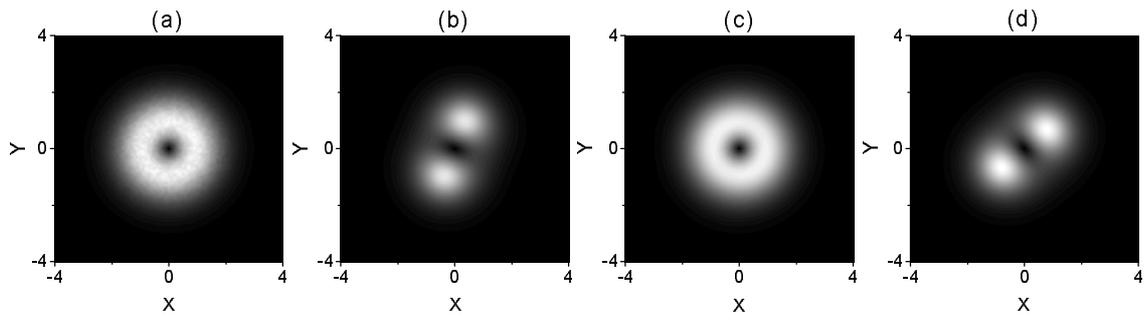
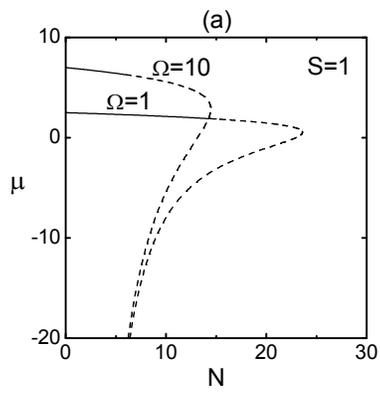
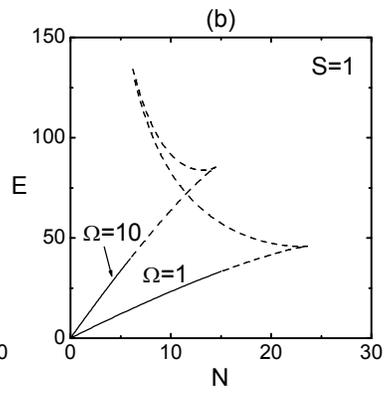
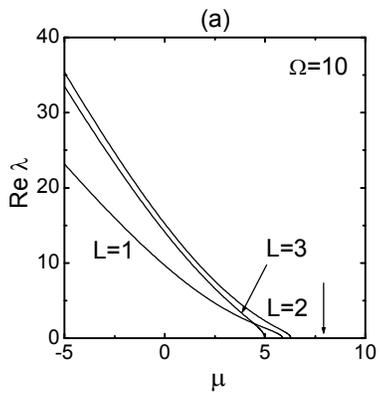
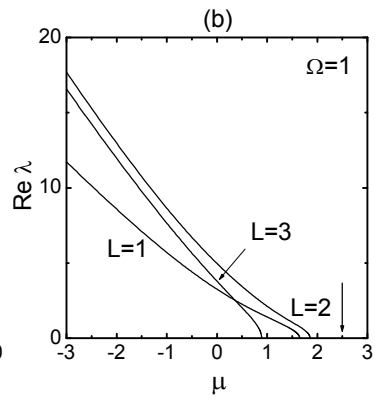
\end{document}